 \documentclass[3p,times]{elsarticle}

\usepackage{amssymb}
 \usepackage{amsmath,bbm}
 \usepackage[figuresright]{rotating}

\def\bfell{\mbox{\boldmath $\ell$}}

\def\La{\Lambda}
\def\Si{\Sigma}

\begin{document}
\begin{frontmatter}

\title{A study of hyperons in nuclear matter based on chiral effective field theory}
\author[J]{J. Haidenbauer}
\author[B,J]{Ulf-G. Mei{\ss}ner}
\address[J]{Institute for Advanced Simulation, Institut f{\"u}r Kernphysik and
J\"ulich Center for Hadron Physics, Forschungszentrum J{\"u}lich, D-52425 J{\"u}lich, Germany}
\address[B]{Helmholtz Institut f\"ur Strahlen- und Kernphysik and Bethe Center
 for Theoretical Physics, Universit\"at Bonn, D-53115 Bonn, Germany}

\begin{abstract}
The in-medium properties of a hyperon-nucleon potential, derived within chiral 
effective field theory and fitted to $\La N$ and $\Si N$ scattering data, are 
investigated. Results for the single-particle potentials of the $\Lambda$ and $\Sigma$ 
hyperons in nuclear matter are reported, based on a conventional $G$-matrix  
calculation. The $\Sigma$-nuclear potential is found to be
repulsive, in agreement with phenomenological information. 
A weak $\Lambda$-nuclear spin-orbit interaction can be achieved by an 
appropriate adjustment of a low-energy constant corresponding to an
antisymmetric $\La N$--$\Si N$ spin-orbit interaction that arises at 
next-to-leading order in the effective field theory approach. 
\end{abstract}
\end{frontmatter}

\section{Introduction}

Chiral effective field theory (EFT) as proposed in the pioneering works of
Weinberg \cite{Wei90,Wei91} is a powerful tool for the derivation of baryonic forces.
In this scheme there is an underlying power counting which allows one to improve calculations
systematically by going to higher orders in a perturbative expansion.
In addition, it is possible to derive two- and corresponding three-body and even four-body forces
as well as external current operators in a consistent way. 
For a review, see e.g. Ref.~\cite{Epelbaum:2008ga}.

Recently, a hyperon-nucleon ($YN$) interaction has been derived up to next-to-leading
order (NLO) in chiral EFT by the J\"ulich-Bonn-Munich group \cite{Hai13}.
At that order there are contributions from one- and two-pseudoscalar-meson exchange 
diagrams (involving $\pi$, $K$, $\eta$) and from four-baryon contact terms without 
and with two derivatives.
SU(3) flavor symmetry is imposed for constructing the $YN$ interaction. The
assumed SU(3) symmetry allows us to fix all the coupling constants at the various
meson-baryon vertices, and it reduces the number of free parameters which 
arise within the EFT approach in form of low-energy constants (LECs) 
associated with the aforementioned contact terms.
In the actual calculation the SU(3) symmetry is broken, however, because
the physical masses of the involved mesons and octet baryons ($N$, 
$\Lambda$, $\Sigma$, $\Xi$) are used. 
With this interaction an excellent reproduction of available low-energy $\Lambda N$ 
and $\Sigma N$ scattering data could be achieved as reflected in a total
$\chi^2$ of about $16$ for the $36$ data points included in the fit \cite{Hai13}.  
Thus, the description of the $YN$ system at NLO is on the same level of quality 
as by the most advanced meson-exchange $YN$ interactions.

In the present work we investigate the properties of our $YN$ interactions in
nuclear matter. Specifically, we report results for the single particle (s.p.)
potential of the $\Lambda$ and $\Sigma$ hyperons in nuclear matter obtained in a 
conventional $G$-matrix calculation based on the standard (gap) choice 
for the intermediate spectrum. Of specific interest for us are two long-standing 
issues that have been discussed extensively in the literature, namely the 
repulsive nature of the $\Sigma$-nucleus potential and the weak 
$\Lambda$-nucleus spin-orbit interaction. For recent overviews of
strangeness nuclear physics see Refs.~\cite{HT06,Gal10,Botta}. 

It is now generally accepted that the $\Sigma$-nuclear potential is repulsive
\cite{Friedman07}. Phenomenological evidence comes, for example, 
from recently measured $(\pi^-,K^+)$ inclusive spectra related to $\Sigma^-$-formation 
in heavy nuclei \cite{SIG2,SIG3} which suggest a repulsive $\Sigma$-nucleus 
potential \cite{Kohno04,Kohno06}. But also the 
theoretical interpretation of measured level shifts 
and widths of $\Sigma^-$ atoms \cite{SIGat1,SIGat2} requires a 
repulsive component of the $\Sigma$-nucleus potential. 
Microscopic models of the $YN$ interaction, fitted to $\La N$ and $\Si N$ scattering
data, often fail to produce a repulsive $\Sigma$-nuclear potential. Specifically,
for models based on meson-exchange dynamics it is rather difficult to obtain such a 
repulsion as is witnessed by results of $G$-matrix calculations 
for various generations of the Nijmegen $YN$ potential
\cite{Rij99,Vid00,Rij06,Rij08}. Only the most recent version \cite{Rij10} 
could overcome this difficulty \cite{Rij10a,Rij13} - but at the expense of 
a noticeable increase in the achieved $\chi^2$.
The situation is different for $YN$ potentials derived within a constituent quark 
model \cite{Fu07}. Here, the interaction in the $^3S_1$ partial wave of the 
$I=3/2$ $\Sigma N$ channel is always strongly repulsive due to Pauli blocking effects on 
the quark level~\cite{Fu07,Ko00}, and as direct consequence of this feature
a repulsive $\Sigma$-nuclear potential is obtained in pertinent $G$-matrix 
calculations.

Interestingly, our leading order (LO) $YN$ interaction, published in 2006 \cite{Po06},
predicted also a repulsive $^3S_1$ partial wave. Calculations of the 
$\Sigma$ s.p. potential, performed by Kohno with a low-energy equivalent 
representation \cite{Ko10}, revealed that this interaction produces indeed a repulsive
$\Sigma$-nuclear potential. The LO interaction involves only five contact terms and
the LECs associated with them were determined in a fit to the 
low-energy $\La N$ and $\Si N$ cross sections. No attention was paid to the sign of
the $I=3/2$ $\Sigma N$ amplitude in the $^3S_1$ partial wave. 
In the course of constructing the EFT interaction \cite{Hai13} up to NLO (which involves
altogether 13 LECs in the $S$ waves and the $S$--$D$ transitions) it turned out
that the available $YN$ scattering data could be fitted equally well with an attractive 
or a repulsive interaction in the $^3S_1$ partial wave of the $I=3/2$ $\Sigma N$ channel.  
For the reasons discussed above, the repulsive solution was adopted. The resulting 
interaction produces a moderately repulsive $^3S_1$ phase shift as can be seen in Fig.~9 
of Ref.~\cite{Hai13}. 
First $G$-matrix calculations for this interaction will be reported in the present
work. It turns out that a repulsive $\Sigma$-nuclear potential is predicted. 

Hypernuclear spectra provide clear indication for a rather weak $\Lambda$-nucleus
spin-orbit potential \cite{HT06,Gal10,Botta}. Apparently, also this feature is difficult
to reproduce with microscopic models of the $YN$ interaction. In particular,
meson-exchange potentials tend to grossly overestimate the spin-orbit force, 
judging from applications of various Nijmegen models in the studies presented in
Refs.~\cite{Hi00} and \cite{Mill10,Mill11}. 
Calculations based on $YN$ interactions derived in the quark model lead to (partially) 
more promising results \cite{Fu07,Fu08}. According to the discussion of the strength 
of the spin-orbit potential in Refs.~\cite{Fu07,Ko00,Fu00}, done on the level of the 
Scheerbaum factor \cite{SCHE} calculated from the $G$-matrix, there is 
a sizeable antisymmetric spin-orbit component generated from the quark model which 
cancels to a large extent the symmetric component in the $\La N$ isospin $I=1/2$ 
channel so that an overall weak spin-orbit potential is achieved. 

The aforementioned work by Kohno \cite{Ko10} provides also results for the 
Scheerbaum factor for our LO $YN$ potential. The predicted value was found to 
be fairly small and -- somewhat surprisingly -- of opposite sign as compared to
results from standard potential models. At leading order there is no contact term
that would give rise to an antisymmetric spin-orbit component. Thus, the result at
LO is a pure prediction based on the dynamics generated by pseudoscalar meson
exchange. 
This is different at NLO where such a contact term occurs and facilitates $^1P_1$--$^3P_1$ 
transitions in the coupled ($I=1/2$) $\La N$--$\Si N$ system \cite{Hai13}. 
Since the corresponding LEC could not be pinned down by a fit to the existing 
$\La N$ and $\Si N$ scattering data it was simply put to zero in Ref.~\cite{Hai13}. 
However, it is interesting to see whether that LEC can be fixed now from a study of
the properties of the $\La$ hyperon in nuclear matter and, specifically, whether
it can be utilized to achieve a weak $\Lambda$-nuclear spin-orbit potential. 
This is indeed the case as will be demonstrated in the present work. 

Note that the same issues have been addressed by Kaiser and 
Weise \cite{KaWe05,Ka05,Ka07,KaWe08} within an approach similar to 
ours, but on a perturbative level. It was found that the iterated (second order)
one-pion exchange (with $\La$ and $\Si$ intermediate states) gives rise to a 
repulsive $\Si$-nuclear potential. Note that dimensional regularization has been 
employed to extract the genuine long-range components. Second-order pion exchange
plays also a decisive role for the spin-orbit force, where it is argued that
its contribution counterbalances the short-range components due to the
scalar and vector mean fields to a large extent in case of the $\La$ and, 
accordingly, leads to a small net result for the $\La$-nuclear spin-orbit force. 
The same mechanism is at work for the $\Sigma$-nuclear spin-orbit coupling 
\cite{Ka07}. An analogous calculation where, in addition, contributions involving
the decuplet baryons $\Delta$ and $\Sigma^*$ were considered \cite{Cam07}
led to the same conclusions concerning the $\La$-nuclear spin-orbit force. 

The paper is structured as follows: 
In Sect.~2 the formalism employed for evaluating the $G$-matrix is outlined. 
Sect.~3 provides a basic introduction to our EFT $YN$ potentials. Specifically, the
implementation of the antisymmetric spin-orbit force and its impact on the
$\La N$ and $\Si N$ results are discussed. 
Results of our calculation are presented and analysed in Sect.~4. Besides
those based on our chiral EFT interactions, predictions for three meson-exchange 
potentials (developed by the J\"ulich and Nijmegen groups, 
respectively) are included for illustration purposes. 
The paper ends with a short summary. 

\section{Formalism for the G-matrix calculation}
The standard formalism for the treatment of a hyperon in nuclear
matter is the conventional Brueckner theory. 
For convenience we summarize below the essential formulae employed in our 
calculation. A more detailed description can be found in Ref.~\cite{Reu94},
see also \cite{Vid00}. 

We consider a $\Lambda$ or $\Sigma$ hyperon with momentum ${\bf p}_Y$ in
nuclear matter of density $\rho = (2/3\pi^2)\,k^3_F$, where $k_F$ denotes
the Fermi momentum of nuclear matter. In order to determine the
properties of those hyperons we employ the Brueckner reaction-matrix
formalism as discussed in Refs.~\cite{B1,B2}. The $YN$ reaction
matrix, $G_{YN}$, is defined by the Bethe-Goldstone equation
\begin{equation}
\langle YN | G_{YN}(\zeta) | YN \rangle = \langle YN | V | YN \rangle 
+\sum_{Y'N} \ \langle YN | V | Y'N \rangle \
\langle Y'N | \frac{Q}{\zeta - H_0}|Y'N \rangle \ \langle Y'N | G_{YN}(\zeta) | YN \rangle , 
\label{Eq:G1}
\end{equation}
with $Y$, $Y'$ = $\La$, $\Si$. 
Here, $Q$ is the Pauli projection operator, which excludes those intermediate
$YN$ states with the nucleon inside the Fermi sea. 
The starting energy $\zeta$ for an initial $YN$ state with momenta ${\bf p}_Y$, ${\bf p}_N$
is given by 
\begin{equation}
\zeta = E_Y (p_Y) + E_N (p_N),
\end{equation} 
where the s.p. energy $E_\alpha (p_\alpha)$ ($\alpha = \Lambda, \Sigma, N$) 
includes not only the (nonrelativistic) kinetic energy and the particle mass but
in addition the s.p. potential $U_\alpha (p_\alpha)$: 
\begin{equation}
E_\alpha (p_\alpha) = M_\alpha + \frac{p^2_\alpha}{2M_\alpha} + U_\alpha (p_\alpha) .
\end{equation} 
Here and in the following, $p_Y$ denotes the modulus of ${\bf p}_Y$, etc. 

In practice, the Bethe-Goldstone equation is solved for a given $Y$ momentum ${\bf p}_Y$
and relative momentum ${\bf p}$ of the initial $\Lambda N$ or $\Sigma N$ state. For
that purpose we re-write the energy denominator in Eq.~(\ref{Eq:G1}) in the form
\begin{eqnarray}
\nonumber
&&e_Y(k;K,\zeta) = \langle {\bf k}_Y,{\bf k}_N | \frac{1}{\zeta - H_0}|{\bf k}_Y,{\bf k}_N \rangle \\
\nonumber
&&= \left(\zeta - M_Y-M_N - \frac{k^2_Y}{2M_Y} - \frac{k^2_N}{2M_N} +{\rm i}\epsilon\right)^{-1} \\
&&= \left(\zeta - M_Y-M_N - \frac{K^2}{2(M_Y+M_N)} - \frac{k^2}{2\mu_{YN}}+{\rm i}\epsilon \right)^{-1},
\label{Eq:3}
\end{eqnarray}
where we have introduced the total and relative momenta, ${\bf K}$ and ${\bf k}$, which are 
related to the $Y$ and $N$ momenta ${\bf k}_Y$ and ${\bf k}_N$ via 
\begin{equation}
{\bf K} = {\bf k}_Y + {\bf k}_N, \qquad {\bf k} = \frac{M_N {\bf k}_Y - M_Y {\bf k}_N}{M_N+M_Y}, 
\end{equation} 
The reduced mass in Eq.~(\ref{Eq:3}) is given by $\mu_{YN} \equiv M_YM_N/(M_Y+M_N)$.

The square of the (conserved) total momentum, ${\bf K}^2$, can be written
as
\begin{equation}
K^2 \equiv {\bf K}^2 = (1+\xi_Y)^2 ({\bf p}_Y - {\bf p})^2 
\end{equation} 
with $\xi_Y = M_N/M_Y$. 
Note that $e_Y$ depends via $K^2$ on the angle between ${\bf p}_Y$ and ${\bf p}$. Also the Pauli 
projection operator, 
\begin{equation}
Q \left(k_N = | -{\bf k} + (1 + \xi^{-1}_{Y'})^{-1} {\bf K}|\right), 
\end{equation}
depends on the angle between the total momentum ${\bf K}$ and the relative momentum
${\bf k}$ of the intermediate $YN$ state. In order to facilitate a partial-wave
decomposition (with regard to the total $YN$ angular momentum ${\bf J}$, with 
${\bf J}$ the sum of the total spin ${\bf S}$ and the relative angular
momentum ${\bf L}$: ${\bf J} = {\bf L}+ {\bf S}$) 
it is common practice to approximate $K^2$ and $Q\,(k_N)$ by their angle-averaged
values, evaluated for given $p_Y$, $p$ of the initial $YN$ state. These are
given by 

\begin{eqnarray}
\nonumber 
\overline{K^2} &=& \overline{K^2}(p_Y,p) \\
&=& \left\{
\begin{array}{l}
({1+\xi_Y})^2(p^2_Y+p^2), \qquad \qquad \rm{for} \quad \xi_Yp_Y+({1+\xi_Y})p \le k_F  \\ 
({1+\xi_Y})^2(p^2_Y+p^2)+\frac{1}{2}(1+\xi^{-1}_Y)[k^2_F-[\xi_Yp_Y+({1+\xi_Y})p]^2], \ \ \rm{otherwise}~, \\ 
\end{array}
\right. 
\end{eqnarray}

\begin{equation}
\overline Q_{Y'}(k;p_Y,p) = \left\{
\begin{array}{l}
0, \qquad \qquad \rm{for} \quad k+{\xi_{Y'} \over 1+\xi_{Y'}}\sqrt{\overline{K^2}} \le k_F  \\ 
1, \qquad \qquad \rm{for} \quad |k-{\xi_{Y'} \over 1+\xi_{Y'}}\sqrt{\overline{K^2}}| > k_F  \\ 
(1+\xi^{-1}_{Y'}) \left[\left(k+{\xi_{Y'} \over 1+\xi_{Y'}}\sqrt{\overline{K^2}}\right)^2 - k^2_F\right]
/ 4k\sqrt{\overline{K^2}}, \ \ \rm{otherwise} ~.\\ 
\end{array}
\right. 
\end{equation} 

When evaluating the starting energy $\zeta = E_Y (p_Y) + E_N (p_N)$ the nucleon 
momentum of the initial $YN$ state, $p_N$, is approximated by 

\begin{equation}
p_N \approx \left[ 
{\xi_Y \over 1+\xi_Y} {\overline K^2} + (1+\xi_Y)p^2 - \xi_Yp^2_Y\right]^{1/2} \ .
\end{equation}

The partial-wave expanded Bethe-Goldstone equation reads      
\begin{eqnarray}
G^{J,I_0}_{\alpha S^\prime L^\prime,\,\alpha S L}(q, p; p_Y)
&=&V^{J,I_0}_{\alpha S^\prime L^\prime,\,\alpha S L}(q,p) \nonumber \\
&+&\sum_{\beta S^{\prime \prime}
L^{\prime \prime}}
 \int^{\infty}_{0} \frac{k^2\,d\,k}{(2\pi)^3} 
\, V^{J,I_0}_{\alpha S^\prime L^\prime,\,\beta S^{\prime \prime}
L^{\prime \prime}}(q, k)
~{{\overline Q}_\beta (k; p_Y,p) e_\beta (k; p_Y,p)}
~G^{J,I_0}_{\beta S^{\prime \prime} L^{\prime \prime},\,\alpha S L}
(k, p; p_Y)\ \ ,
\label{jh01}
\end{eqnarray}
where $\alpha$ stands for $\Lambda$ or $\Sigma$ and $I_0$ is the total $YN$ isospin
which can be $1/2$ or $3/2$. 

The hyperon s.p. potential $U_Y$ is calculated from
\begin{eqnarray}
& & U_Y(p_Y)=(1+\xi_Y)^3 \sum_{I_0} {2I_0+1 \over 2(2I_Y+1)} 
\, \sum_{JSL} (2J+1) 
\int_{0}^{p_{max}} \frac{p^2\,d\,p}{(2\pi)^3}
~W(p, p_Y)~G^{J,I_0}_{YSL,YSL}(p,p; p_Y)\ \ ,
\label{jh02}
\end{eqnarray}
where $p_{max}=(k_F+\xi_Y p_Y)/(1+\xi_Y)$.
The weight function $W(p, p_Y)$ is given by
\begin{equation}
W(p,p_Y) = \left\{
\begin{array}{l}
1, \qquad \qquad \rm{for} \quad p \le {k_F - \xi_Yp_Y \over 1+\xi_Y}, \\
0, \qquad \qquad \rm{for} \quad |\,\xi_Yp_Y- (1+\xi_Y)p\,| > k_F, \\
{k^2_F - [\xi_Yp_Y - (1+\xi_Y)p]^2 \over 4\xi_Y (1+\xi_Y)p_Y p}, \ \ 
\rm{otherwise}~. \\ 
\end{array}
\right. 
\end{equation} 
We evaluate the $\Lambda$ and $\Sigma$ s.p. potentials $U_\La$ and $U_\Si$ 
self-consistently in the standard way for a fixed Fermi momentum $k_F$. 
In case of the $\Si$ hyperon we decompose the s.p. potential into
an isoscalar and an isovector based on the parameterization \cite{Gal10} 
\begin{equation}
U_\Si = U_\Si^0 + \frac{1}{A} U_\Si^1 \ {\bf t}_\Si \cdot{\bf T}_A \ .
\label{iso}
\end{equation} 
Here, ${\bf t}_\Si$ and ${\bf T}_A$ are the isospin operators of the
$\Sigma$ and the nucleus (with $A$ nucleons). The isoscalar and isovector 
components are given by $U^0_\Si = U_{\frac{3}{2}} + U_{\frac{1}{2}}$ and
$U^1_\Si = U_{\frac{3}{2}} - 2 U_{\frac{1}{2}}$, respectively, where $U_{I_0}$ 
are the corresponding parts of the sum over $I_0$ in Eq.~(\ref{jh02}).

In case of the $\Si$ hyperon, the propagator in Eq.~(\ref{Eq:3}) can become singular if 
the $\Si$ s.p. potential is repulsive. We avoided the proper but technically involved 
treatment of this singularity by adding a small finite imaginary part in our calculation. 
We verified that this prescription has no influence on the obtained results for the real 
part of the $\Si$ s.p. potential. However, we cannot calculate the imaginary part of $U_\Si$. 
In any case, it is known that the imaginary part can be only determined reliably when
using the continuous choice for the intermediate spectrum \cite{Rij99} and we plan to
perform such a calculation in the future \cite{Pet15}. 
In the present work (like before in \cite{Reu94}), the nucleon s.p. potential $U_N$ is
taken from a calculation \cite{Yamamoto} of pure nuclear matter employing a phenomenological
$NN$ potential. It has been found that the results for the $\Lambda$ and $\Sigma$
hyperon in nuclear matter for the gap choice are not too sensitive to the choice 
of $U_N$ \cite{Reu94}. Nonetheless, also this limitation will be avoided in the 
future calculation \cite{Pet15}. 

In order to relate the strength of the $\Lambda$-nuclear spin-orbit potential 
to the two-body $\Lambda N$ interaction we consider the Scheerbaum factor 
$S_Y$ \cite{SCHE} calculated in nuclear matter. In the Scheerbaum approximation, 
the $\Lambda$- or $\Sigma$-nuclear spin-orbit potential is represented by
\begin{equation}
 U_{Y}^{\ell s}(r)=-\frac{\pi}{2} S_Y \frac{1}{r} 
 \frac{d\rho (r)}{dr}\mbox{\boldmath{$\ell$}}\cdot\mbox{\boldmath{$\sigma$}},
\label{SB1}
\end{equation}
with $\rho (r)$ the nucleon density distribution and ${\bfell}$ the 
single-particle orbital angular momentum operator. We adopt here the
definition as given in Ref.~\cite{Fu00}. The explicit expression for the 
Scheerbaum factor $S_Y$ derived in this reference in terms of the $G$-matrix 
elements given in momentum space reads in our notation 
\begin{eqnarray}
& & S_Y(p_Y) = -
~{3\pi \over 4(k_F)^3}~\xi_Y(1+\xi_Y)^2
\sum_{I_0,J} {2I_0+1 \over (2I_Y +1)} (2J+1)\,
\int_0^{p_{max}} \frac{d\,p}{(2\pi)^3}~W(p, p_Y)
~\left\{~(J+2)\,G^{J,I_0}_{Y1\,J+1,\,Y1\,J+1}(p, p; p_Y)
\right. \nonumber \\ 
& & +\,G^{J,I_0}_{Y1J, Y1J}(p, p; p_Y)
-(J-1) \,G^{J,I_0}_{Y1\,J-1,\,Y1\,J-1}(p, p; p_Y) 
 \left. -\sqrt{J(J+1)}~\left[~G^{J,I_0}_{Y1J, Y0J}(p, p; p_Y)
+G^{J,I_0}_{Y0J, Y1J}(p, p; p_Y) \right] \right\}\ \ .\nonumber \\ 
\label{SB2}
\end{eqnarray}

\section{The hyperon-nucleon interaction in chiral EFT}

A comprehensive description of the derivation of the chiral baryon-baryon 
potentials for the strangeness sector using the Weinberg power counting can be 
found in Refs.~\cite{Hai13,Po06,Haidenbauer:2007ra,Pet13}.
The LO potential consists of four-baryon contact terms without derivatives and of
one-pseudoscalar-meson exchanges while at NLO contact terms with two derivatives
arise, together with contributions from (irreducible) two-pseudoscalar-meson exchanges.
The contributions from pseudoscalar-meson exchanges ($\pi$, $\eta$, $K$) are completely
fixed by the assumed SU(3) flavor symmetry in terms of the $F$ and $D$ couplings.
On the other hand, the strength parameters associated with the contact terms, the LECs,
need to be determined by a fit to data. How this is done is described in detail in
Ref.~\cite{Hai13} and, therefore, we will be brief here. Since SU(3) symmetry is
also imposed for the contact terms, the number of independent LECs that can contribute 
is significantly reduced. In particular, for the $\La N$--$\Si N$ system there 
are in total 13 LECs entering the $S$-waves and the $S$--$D$ transitions, respectively,
and 10 constants in the $P$-waves. The $S$-wave LECs could be fixed by a fit to 
available low-energy total cross sections for the reactions $\Lambda p \to \Lambda p$,
$\Sigma^-p \to \Lambda n$, $\Sigma^-p \to \Sigma^0 n$, $\Sigma^-p \to \Sigma^- p$, 
and $\Sigma^+p \to \Sigma^+ p$. Information from the nucleon-nucleon ($NN$) 
sector was not needed.
The limited number (and quality) of differential cross sections and the complete
lack of polarization observables makes a determination of the contact terms in
the $P$-waves from $\La N$--$\Si N$ data alone impossible. 
Therefore, in this case $NN$ $P$-wave phase shifts had to be used as a further 
constraint for the contact terms. 
The LEC that generates the antisymmetric spin-orbit force, and which produces 
a $^1P_1$-$^3P_1$ transition, was set to zero because 
it could not be determined from the $YN$ scattering data.

The reaction amplitudes are obtained from the solution of a coupled-channel
Lippmann-Schwinger (LS) equation for the interaction potentials \cite{Hai13}.
The potentials in the LS equation are cut off with a regulator function, $f_R(\Lambda) =
\exp\left[-\left(p'^4+p^4\right)/\Lambda^4\right]$,
in order to remove high-energy components \cite{Epe05}.
In Ref.~\cite{Hai13} cutoff values in the range $\Lambda=450$ -- $700\,$MeV are considered, 
similar to what was used for chiral $NN$ potentials \cite{Epe05}. The variation of the
results with the cutoff can be viewed as a rough estimate for the theoretical 
uncertainty \cite{Epe05}.

As demonstrated in Ref.~\cite{Hai13}, the available $\Lambda N$ and $\Sigma N$ scattering 
data are very well described by the EFT interaction. Indeed the
description of the $YN$ system achieved at NLO is on the same level of
quality as the one by the most advanced meson-exchange $YN$ interactions.
Besides an excellent description of the $YN$ data the chiral EFT interaction yields
a satisfactory value for the hypertriton binding energy, see Ref.~\cite{Hai13}.
Calculations for the four-body hypernuclei ${}^4_\Lambda {\rm H}$ and ${}^4_\Lambda {\rm He}$
based on the EFT interactions can be found in Ref.~\cite{Nog14}.

Let us come back to the antisymmetric spin-orbit force. 
The general structure of the spin- and momentum-dependence of the contact terms 
at next-to-leading order is given by \cite{Hai13} 
\begin{eqnarray}
V_{YN\to YN} &=& C_1 {\bf q}^{\,2}+ C_2 {\bf k}^{\,2} + (C_3 {\bf q}^{\,2}+ C_4 {\bf k}^{\,2})
\,\mbox{\boldmath $\sigma$}_1\cdot\mbox{\boldmath $\sigma$}_2 
+ \frac{i}{2} C_5 (\mbox{\boldmath $\sigma$}_1+\mbox{\boldmath $\sigma$}_2)\cdot ({\bf q} \times {\bf k}) \nonumber \\
&+& C_6 ({\bf q} \cdot \mbox{\boldmath $\sigma$}_1) ({\bf q} \cdot \mbox{\boldmath $\sigma$}_2)
+ C_7 ({\bf k} \cdot \mbox{\boldmath $\sigma$}_1) ({\bf k} \cdot \mbox{\boldmath $\sigma$}_2)
+ \frac{i}{2} C_8 (\mbox{\boldmath $\sigma$}_1-\mbox{\boldmath $\sigma$}_2)\cdot ({\bf q} \times {\bf k}) \ , 
\label{VC}
\end{eqnarray}
where  $C_i$ ($i=1,\dots,8$) are the LECs. 
The transferred and average momenta, ${\bf q}$ and ${\bf k}$, are defined in terms of
the final and initial center-of-mass momenta of the baryons, ${\bf p}'$ and ${\bf p}$, as
${\bf q}={\bf p}'-{\bf p}$ and ${\bf k}=({\bf p}'+{\bf p})/2$.

The term proportional to $C_8$ in Eq.~(\ref{VC}) represents an antisymmetric spin-orbit 
force and gives rise to (spin) singlet-triplet ($^1P_1 - {^3P_1}$) transitions.
After performing a partial-wave projection and imposing SU(3) symmetry constraints
the $^1P_1$-$^3P_1$ transition potentials read in the notation of Ref.~\cite{Hai13}
\begin{eqnarray}
\nonumber
V_{\La N \to \La N}^{I=1/2}(^1P_1 \to {^3P_1}) &=& -{C}^{8_s8_a}\, {p}\, {p}'~, \\
\nonumber
V_{\La N \to \Si N}^{I=1/2}(^1P_1 \to {^3P_1}) &=& -3{C}^{8_s8_a}\, {p}\, {p}'~, \\
\nonumber
V_{\La N \to \Si N}^{I=1/2}(^3P_1 \to {^1P_1}) &=& \phantom{-}{C}^{8_s8_a}\, {p}\, {p}'~, \\
\nonumber
V_{\Si N \to \Si N}^{I=1/2}(^1P_1 \to {^3P_1}) &=& \phantom{-}3{C}^{8_s8_a}\, {p}\, {p}'~, \\
V_{\Si N \to \Si N}^{I=1/3}(^1P_1 \to {^3P_1}) &=& \phantom{-}0~,
\label{VC1}
\end{eqnarray}
with $p = |{\bf p}\,|$ and ${p}' = |{\bf p}\,'|$. 
Expressions for the time-reversed transitions follow from the requirement of time-reversal
invariance. The superscript ${8_s8_a}$ of the
LEC is a reminder that the singlet-triplet transition is accompanied by a transition 
between the symmetric and antisymmetric SU(3) octet representations as required by the 
generalized Pauli principle. Note that, in line with Ref.~\cite{Pet13},
we have absorbed here the overall factor $1/\sqrt{20}$ (see Table 1 in 
Ref.~\cite{Hai13}) into the definition of ${C}^{8_s8_a}$. 

In the present work we fix the LEC ${C}^{8_s8_a}$ by considering the Scheerbaum factor 
$S_\Lambda$ \cite{SCHE} calculated from the $G$-matrix. With regard to its ``empirical'' 
value we took studies of the splitting of the $5/2$ and $3/2$ states of 
$^{9}_\Lambda {\rm Be}$ by Hiyama et al.~\cite{Hi00} and Fujiwara et 
al.~\cite{Fu04} as guideline. The results of those authors suggest that values for 
$S_\Lambda$ in the order of $-4.6$ to $-3.0$ MeV~fm$^5$ would be needed to reproduce 
the experimentally observed small level splitting \cite{Ko14}. Values of around $-3.2$ 
and $-4.1$ MeV~fm$^5$ are suggested in Refs.~\cite{Ko10} and \cite{Fu08}, respectively. 
Since the precise value required for the $\Lambda$ s.p. spin-orbit strength can  only be 
pinned down via a dedicated calculation of finite hypernuclei based on our EFT interactions,
which is beyond the scope of the present study, we decided to aim at a typical
result of $-3.7$ MeV~fm$^5$. Indeed, any reference value at the expected order 
of magnitude would be sufficient for the main goal of the present investigation, namely for 
exploring in how far the LEC ${C}^{8_s8_a}$ can be used to achieve a small $\Lambda$ s.p. 
spin-orbit potential.

\begin{figure}[t]
\begin{center}
\includegraphics[height=64mm]{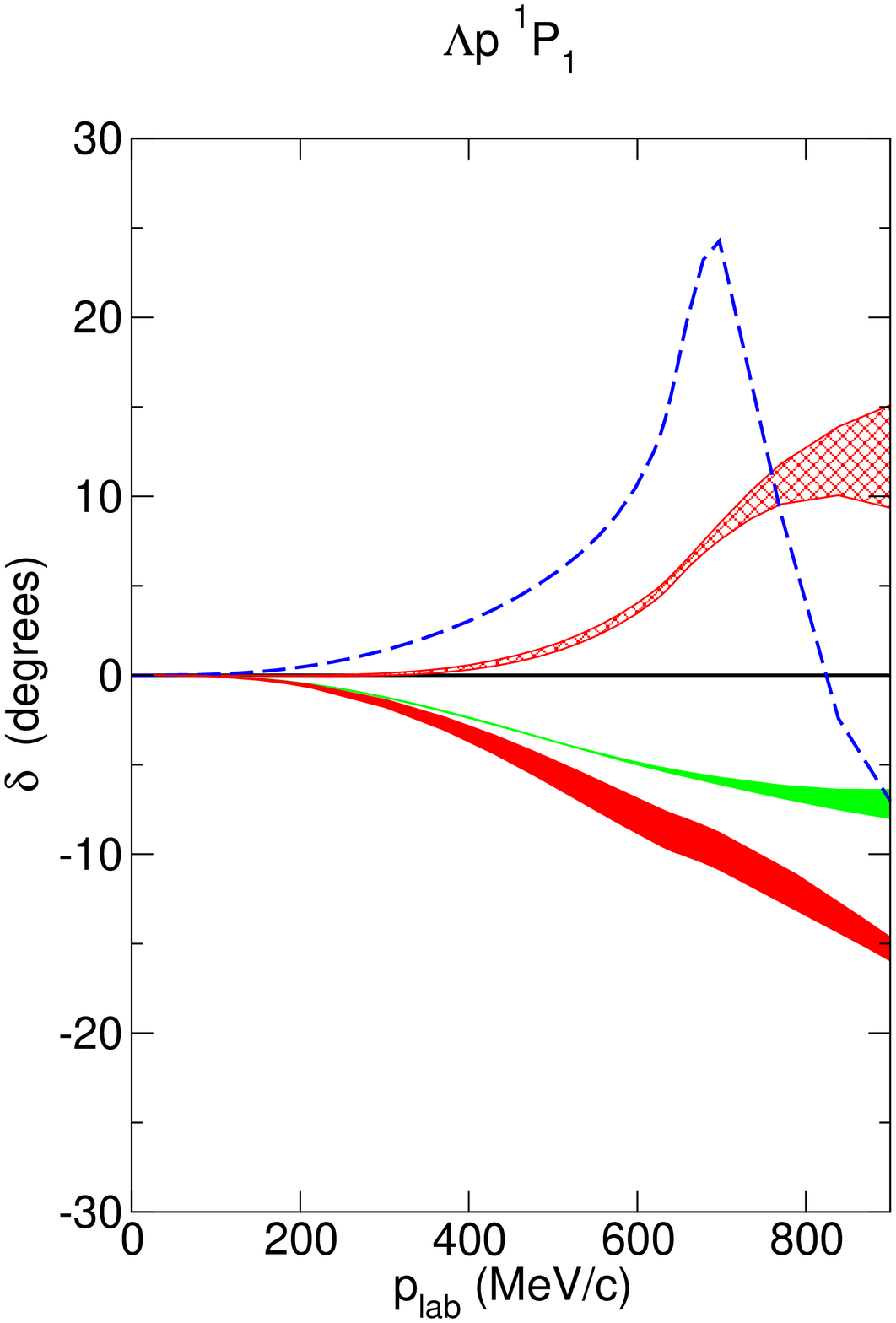}~~~~~
\includegraphics[height=64mm]{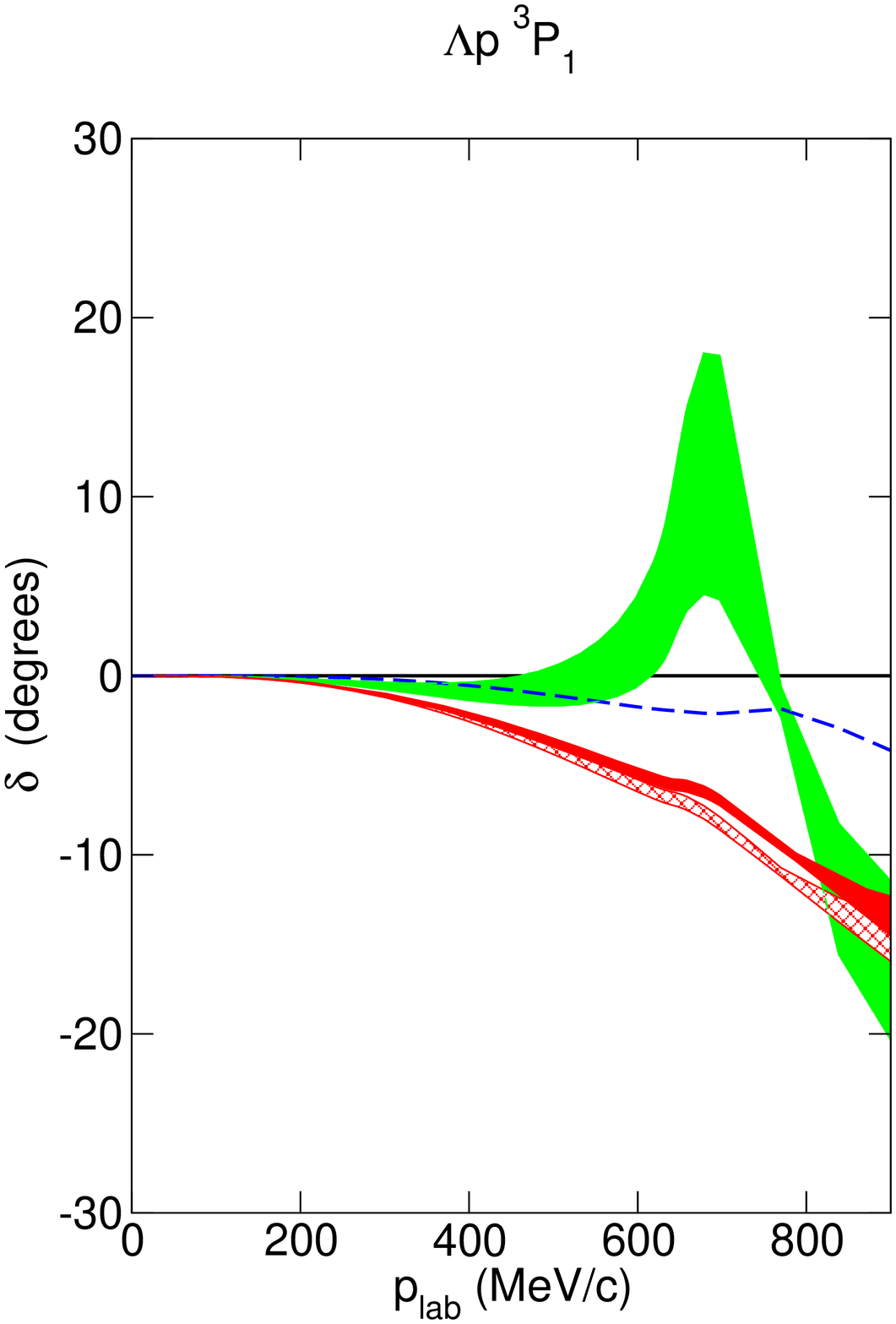}~~~~~
\includegraphics[height=64mm]{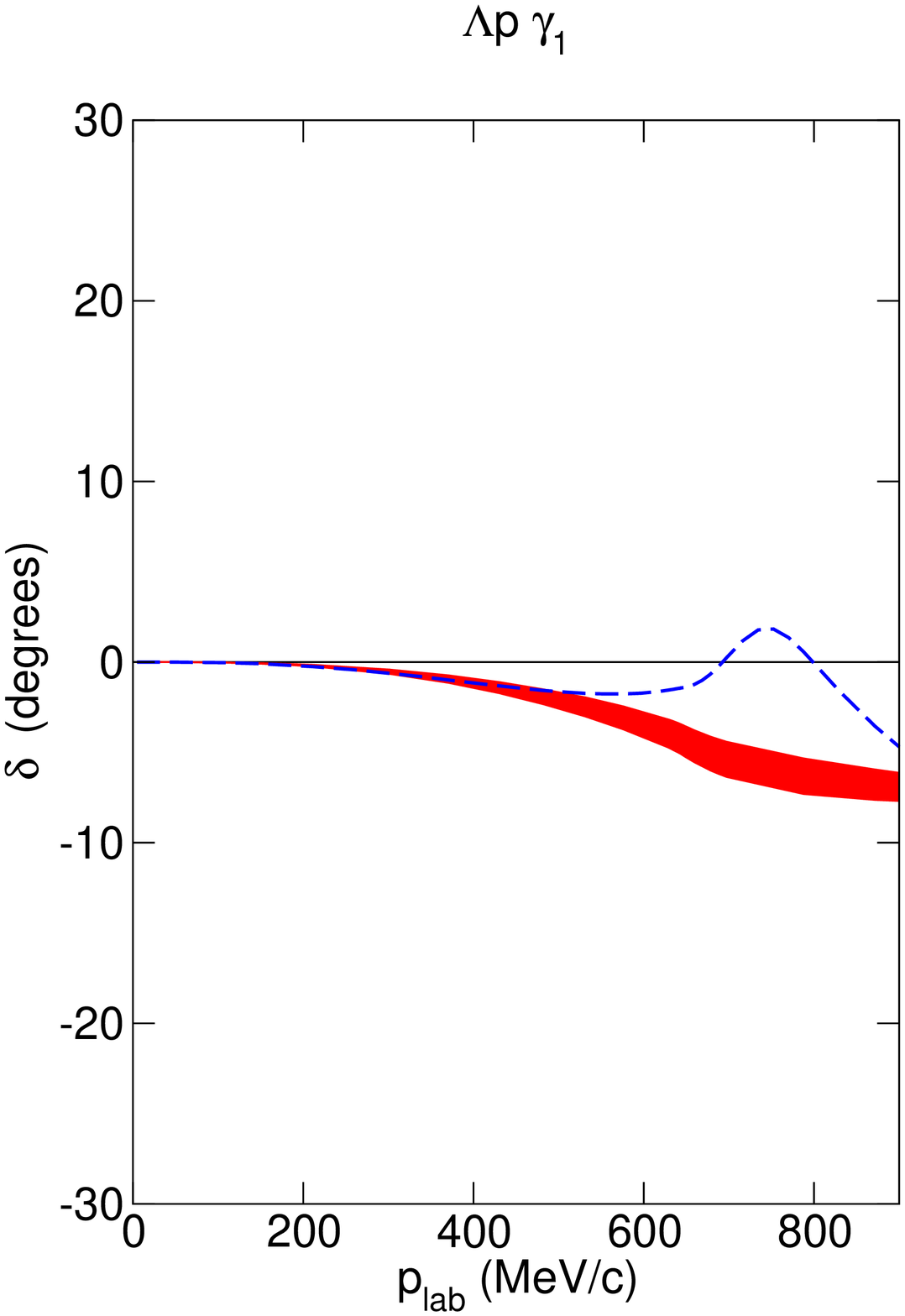}
\caption{
The $\Lambda p$ $^1P_1$ and $^3P_1$  phase shifts and the $^1P_1$--$^3P_1$ mixing parameter $\gamma_1$
as a function of $p_{{\rm lab}}$.
The red/dark band shows the chiral EFT results to NLO for variations of the cutoff
in the range $\Lambda =$ 500,$\ldots$,650~MeV,
while the green/light band are results to LO for $\Lambda =$ 550,$\ldots$,700~MeV.
The hatched band is the NLO fit from \cite{Hai13} without antisymmetric spin-orbit force. 
The dashed curve is the result of the J{\"u}lich '04 meson-exchange potential \cite{Hai05}.
}
\label{fig:Phase}
\end{center}
\end{figure}

\begin{table}
\caption{The $YN$ contact terms for the $^1P_1$ partial wave and the
$^1P_1$--$^3P_1$ transition potential for various cut--offs.
The values of the LECs are in
$10^4$ ${\rm GeV}^{-4}$, the values of $\Lambda$ in MeV.
}
\renewcommand{\arraystretch}{1.2}
\label{tab:F1}
\vspace{0.2cm}
\centering
\begin{tabular}{|c|c|rrrrrr|}
\hline
\multicolumn{2}{|c|}{$\Lambda$} & $450$ & $500$ & $550$& $600$& $650$& $700$  \\
\hline
$^1P_1$
&$C^{10}_{^1P_1}$  &$0.49$ &$0.49$  &$0.49$ &$0.49$  &$0.49$  &$0.49$\\
&$C^{10^*}_{^1P_1}$&$-0.14$ &$-0.14$  &$-0.14$ &$-0.14$  &$-0.14$  &$-0.14$\\
&$C^{8_a}_{^1P_1}$ &$-0.05$ &$-0.15$  &$-0.18$ &$-0.21$  &$-0.23$  &$-0.25$\\
\hline
$^1P_1$-$^3P_1$
&$C^{8_s8_a}$& $-0.084$ &$-0.073$  &$-0.065$ &$-0.059$  &$-0.053$  &$-0.048$\\
\hline
\end{tabular}
\renewcommand{\arraystretch}{1.0}
\label{tab:LEC}
\end{table}

Of course, while introducing a non-zero contact term for the antisymmetric spin-orbit force 
into our EFT interaction, we still want to maintain the excellent description of the $\La N$ and 
$\Si N$ scattering data as reflected in the total $\chi^2$ of only $16$ -- $17$ for the $36$ data 
points considered in Ref.~\cite{Hai13}. It turned out that this is easily possible and no refit
of the $S$-wave LECs is required. However, we had to re-adjust some LECs in the $^1P_1$ 
partial wave in order to preserve the reproduction of the trend shown by the 
$\Si^- p \to \La n$ differential cross sections at $p_{\rm lab} = 135 - 160$ MeV/c, 
cf. Fig.~4 of Ref.~\cite{Hai13}. This resulted in a change of the sign of the
predicted $\La p$ $^1P_1$ phase shift. Corresponding results are shown in 
Fig.~\ref{fig:Phase} where also the $^1P_1$--$^3P_1$ mixing parameter $\gamma_1$ 
is presented. 

The relevant LECs are summarized in Table~\ref{tab:LEC}. All other LECs are kept as given in
Table~3 of Ref.~\cite{Hai13}. In this context let us mention, however, that unfortunately
there is a typo in the latter table. The values for $\tilde C^{10}_{^3S_1}$ should read
0.104, 0.541, 1.49, 3.44, 4.99, and 5.60 for increasing cutoffs, i.e. the comma was misplaced
in case of the last 4 entries. In the following we discuss NLO results for cutoffs in the
range of 500--650~MeV for which the best results for $\La N$ and $\Si N$ scattering were
achieved \cite{Hai13}. 


\section{Results}

Let us now discuss the properties of our $YN$ interactions in nuclear matter.
As already said, we performed a conventional $G$-matrix calculation
based on the standard (gap) choice for the intermediate spectrum. 
Table~\ref{tab:La} summarizes the results for the $\Lambda$ potential depth,
$U_\Lambda (p_\Lambda = 0)$, evaluated at the saturation
point of nuclear matter, i.e. for $k_F=1.35$~fm$^{-1}$, obtained from our LO 
and NLO EFT interactions for the considered cutoff range.
For illustration we include also exemplary results based on the $YN$ 
interaction as published in \cite{Hai13}, i.e. without antisymmetric spin-orbit 
force, and, in addition, 
of a $YN$ interaction where all two-meson exchange contributions that involve 
the heavy mesons $\eta$ and/or $K$ were omitted and only the $\pi\pi$ exchange diagrams 
were kept, which was also considered in Ref.~\cite{Hai13}. 
In this case only results for a single cutoff ($\Lambda = 650$~MeV and 
$\Lambda = 600$~MeV, respectively) are presented. 
Results obtained for the J\"ulich meson-exchange potentials from 2004 \cite{Hai05} and 
1994 (model $\tilde A$) \cite{Reu94} and for the Nijmegen NSC97f model \cite{Rij99} are 
also presented. 
Some in-medium results for the J\"ulich potentials have been published 
before \cite{Reu94,Samma,Hu14}.

As can be seen from the contributions to the various partial waves, there is a 
moderate cutoff dependence at LO and NLO. Indeed, the variations at NLO are 
slightly larger than those at LO -- contrary to the trend we observed for
$YN$ scattering in free space \cite{Hai13} and for the binding energies of
light hypernuclei \cite{Nog14}. 
\begin{table}[h]
\renewcommand{\arraystretch}{1.2}
\centering
\caption{
{Partial-wave contributions} {to} $ U_\Lambda (p_\Lambda = 0)$ (in MeV)
{at} $k_F = 1.35 \ {\rm fm}^{-1}$ for the LO and NLO interactions for various
cutoffs. 
NLO$^\dagger$(650) are results based on the interaction as published in Ref. \cite{Hai13}.
NLO (600$^*$) denotes result for an interaction where all two-meson-exchange contributions
involving the $\eta$- and/or $K$ meson have been omitted, cf. text.
}
\begin{tabular}{|c|rcrrrc|r|}
\hline
& $^1S_0$ & $^3S_1+^3D_1$ & $^3P_0$ & $^1P_1$ & $^3P_1$ & $^3P_2+^3F_2$ & Total \\
\hline
LO (550)  &  $-$12.5 &  $-$26.6 &   $-$1.6 &    1.5 &    1.8 &   $-$0.3 &  $-$38.0 \\
LO (600)  &  $-$12.0 &  $-$25.4 &   $-$1.7 &    1.5 &    1.8 &   $-$0.4 &  $-$36.5 \\
LO (650)  &  $-$11.6 &  $-$24.3 &   $-$1.8 &    1.5 &    1.7 &   $-$0.4 &  $-$35.2 \\
LO (700)  &  $-$11.6 &  $-$23.1 &   $-$1.9 &    1.5 &    1.6 &   $-$0.5 &  $-$34.4 \\
\hline
NLO (500)  &  $-$15.3 &  $-$15.8 &    1.1 &    2.3 &  1.1 &   $-$1.3 &  $-$28.2  \\
NLO (550)  &  $-$13.9 &  $-$12.5 &    1.0 &    2.1 &  1.0 &   $-$1.2 &  $-$23.8  \\
NLO (600)  &  $-$12.6 &  $-$12.0 &    0.9 &    1.9 &  0.8 &   $-$1.1 &  $-$22.4  \\
NLO (650)  &  $-$11.6 &  $-$13.4 &    0.8 &    1.8 &  0.7 &   $-$1.1 &  $-$23.2  \\
\hline
NLO$^\dagger$(650) &  $-$11.6 &  $-$13.4 &    0.8 &    0.4 &    1.6 &   $-$1.1 &  $-$23.5 \\
NLO (600$^*$) &  $-$12.5 &  $-$15.9 &   $-$0.1 &    1.1 &    1.1 &    0.6 &  $-$26.0 \\
{J\"ulich '04}  & $-$10.2 & $-$36.3 & $-$0.7 & $-$0.6 &   0.5 & $-$3.2 & $-$51.2 \\
{J\"ulich '94}  & $-$3.6 & $-$27.2 & 0.6 &   0.7 & 1.4 & $-$0.8 & $-$29.8 \\
NSC97f          & $-$14.6 & $-$23.1 & 0.5 &   2.4 &  4.6 & $-$0.8 & $-$32.4 \\
\hline
\end{tabular}
\label{tab:La} 
\renewcommand{\arraystretch}{1.0}
\end{table}
Comparing our results at NLO with those at LO one notices that the former provides
more attraction in the $^1S_0$ partial wave but significantly less in the coupled
$^3S_1$--$^3D_1$ partial wave. Indeed the contribution predicted at NLO for this 
state is also much smaller than the values suggested by the J\"ulich meson-exchange 
potentials and other phenomenological models \cite{Rij99,Rij06,Rij08,Fu07}. 
We believe that this has to do with a characteristic feature of our EFT 
interactions that was already emphasized in Ref.~\cite{Hai13}: 
The $\Lambda p$ $T$-matrix for the $^3S_1$$\leftrightarrow$$^3D_1$ transition is
rather large and, specifically, of the same magnitude as the ones for 
$^3S_1$$\to$$^3S_1$ and $^3D_1$$\to$$^3D_1$ at energies around the $\Si N$
threshold. 
While all three amplitudes contribute to the $\Lambda p$ cross section,
the corresponding off-diagonal element of the $G$-matrix does not enter
into the evaluation of $U_\Lambda$, see Eq.~(\ref{jh02}). 
The $P$-wave contributions generated by the LO and NLO interactions are of 
comparable magnitude. However, the $^3P_0$ contribution changes sign and 
is repulsive in the NLO case. Note that at LO there are no contact terms
in the $P$-waves and, therefore, their contributions are determined solely 
by one-meson exchange ($\pi$, $K$, $\eta$). The partial wave contributions
obtained for the NLO interaction of Ref.~\cite{Hai13} are practically 
identical to those predicted by the present interaction that includes an
antisymmetric spin-orbit force, cf. NLO$^\dagger$(650) versus NLO (650) in Table
\ref{tab:La}, with the exception that the strengths provided by the $^1P_1$ 
and $^3P_1$ partial waves are interchanged. 

In lowest order of the hole-line expansion $U_\Lambda (p_\Lambda = 0)$ has
to be compared with the 'empirical' value for the $\Lambda$ binding energy in 
nuclear matter of about $-28$ MeV, deduced from the binding energies of 
finite $\Lambda$ hypernuclei \cite{Millener88,Yamamoto88}. The results for
our EFT interactions, given in the last column of Table~\ref{tab:La}, are in 
good qualitative agreement with this value. One should keep in
mind that, at this order of the hole-line expansion, the result depends somewhat 
on the choice of the single-particle potentials, see Ref.~\cite{Rij99} for example, 
so that only a comparison on a qualitative level is meaningful. 

\begin{figure}[t]
\begin{center}
\includegraphics[height=84mm]{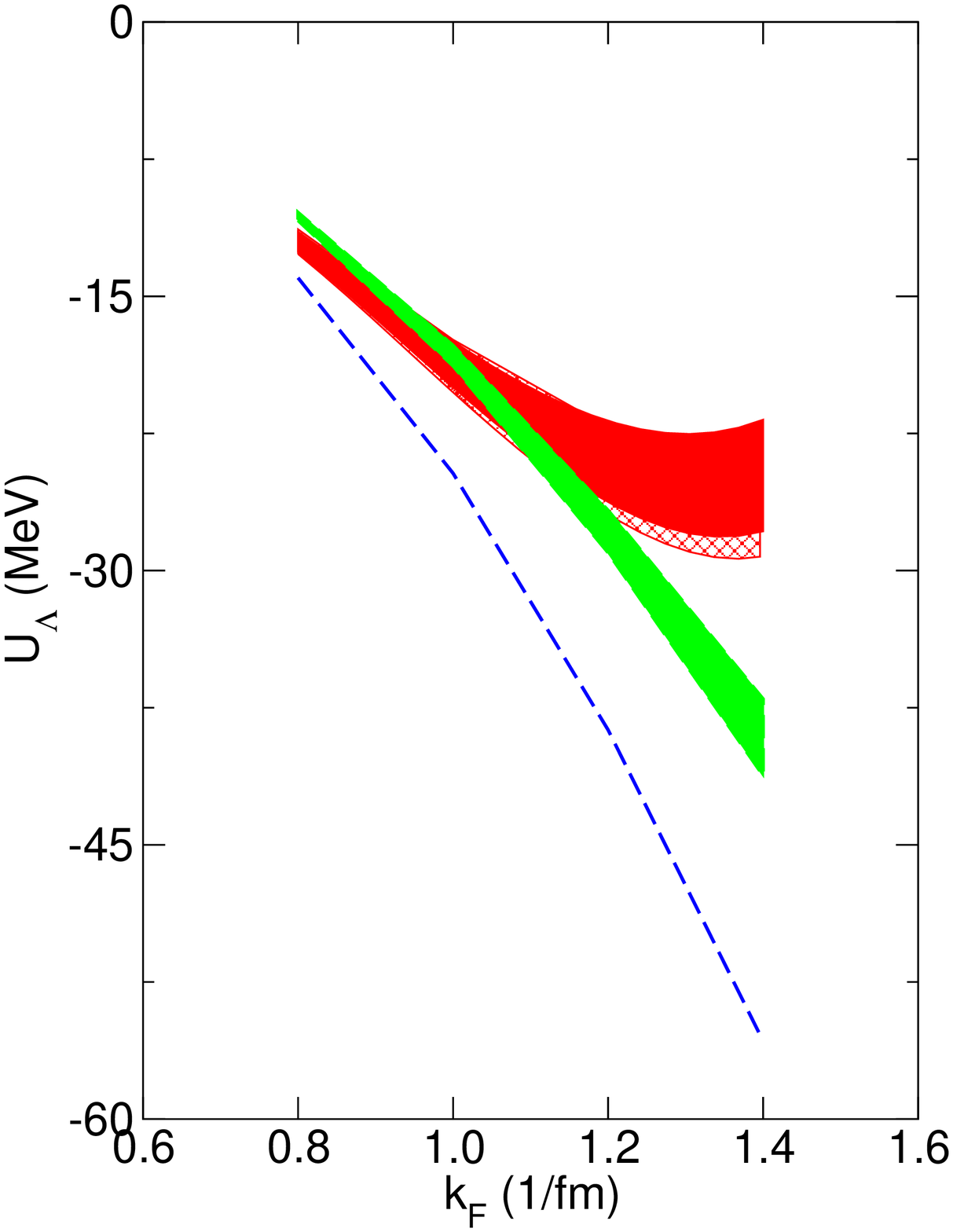}~~~~~~
\includegraphics[height=84mm]{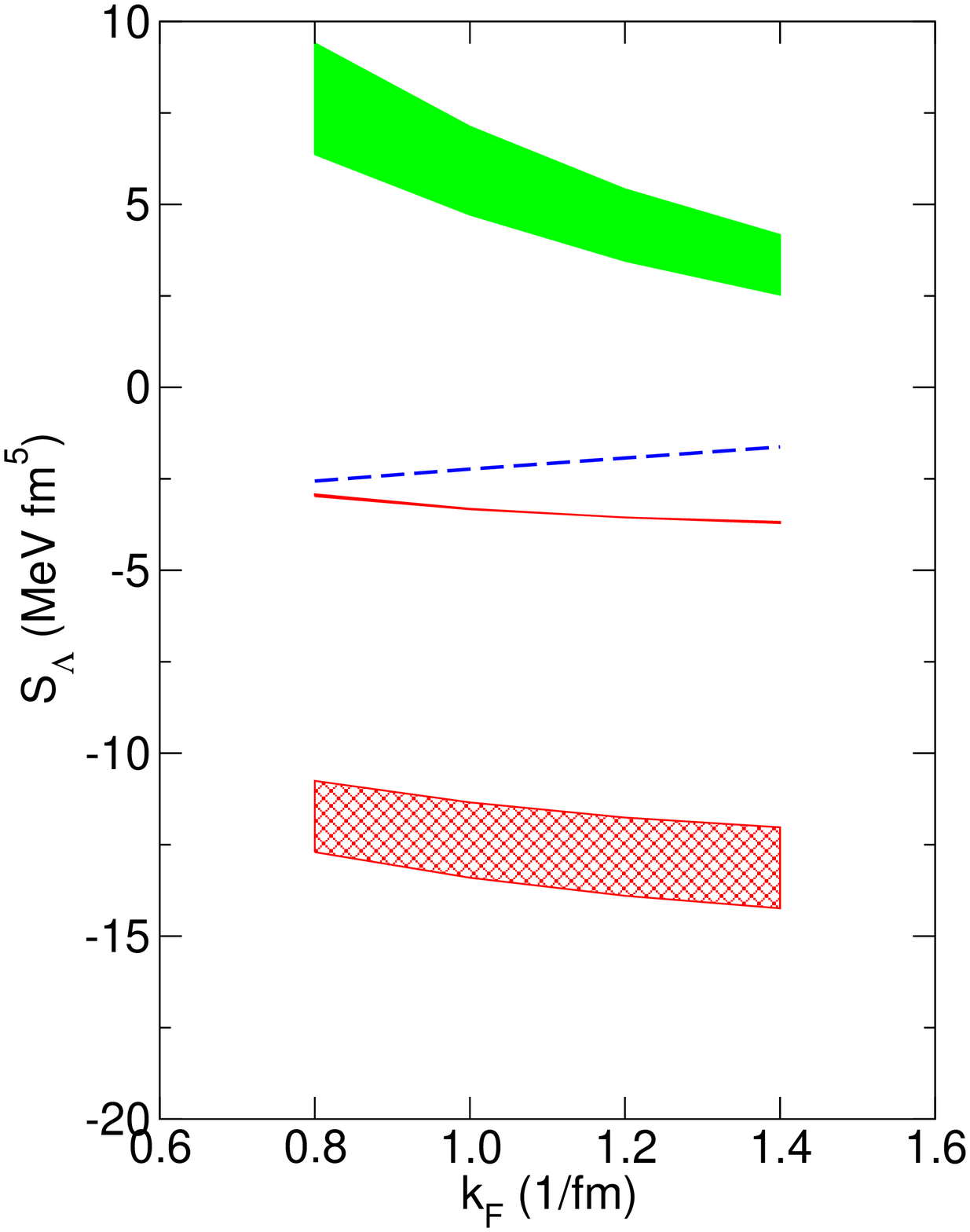}
\caption{The $\Lambda$ s.p. potential $U_\Lambda (p_\Lambda = 0)$ (left) and
the Scheerbaum factor $S_\La$ (right), as a function of the Fermi momentum $k_F$.
The red/dark band shows the chiral EFT results to NLO for variations of the cutoff
in the range $\Lambda =$ 500,$\ldots$,650~MeV,
while the green/light band are results to LO for $\Lambda =$ 550,$\ldots$,700~MeV.
The hatched band is the NLO fit from \cite{Hai13} without antisymmetric spin-orbit force. 
The dashed curve is the result of the J{\"u}lich '04 meson-exchange potential \cite{Hai05}.
}
\label{fig:U}
\end{center}
\end{figure}

Predictions for the density dependence of the $\Lambda$ s.p. potential are presented 
in Fig.~\ref{fig:U} where $U_\Lambda (p_\Lambda = 0)$ is shown as a function of $k_F$ 
for the various interactions. Obviously the EFT interaction at NLO exhibits a
relatively weak density dependence as compared to that of the LO interaction and also to
the J\"ulich '04 potential. Again, most likely this is due to the strong tensor coupling 
in the $^3S_1$--$^3D_1$ partial wave in combination with the $\La N$--$\Si N$ coupling 
already discussed above. Interestingly, one observes an onset of repulsive effects already 
at rather moderate densities. Other potentials, like the NSC97f model, exhibit such a 
behavior too but only at somewhat higher densities \cite{Rij06}. 
The result based on the NLO interaction from \cite{Hai13} without antisymmetric spin-orbit
component coincides more or less with the present one with an antisymmetric spin-orbit
force. 

Results for the $\Lambda$-nuclear spin-orbit potential, given in form of the Scheerbaum 
factor $S_\La$ calculated in nuclear matter, see Eqs.~(\ref{SB1})-(\ref{SB2}), are
summarized in Table~\ref{tab:S}. Besides the total $S_\La$  we list also the 
contributions of the individual partial waves.
As already said above, it has been established experimentally that the 
spin-orbit splitting of the $\Lambda$ s.p. levels in nuclei is very 
small \cite{HT06}. Accordingly, the $\Lambda$ s.p. spin-orbit potential
should also be small. In terms of the Scheerbaum factor, the value
of $S_\Lambda$ required for explaining experimental data is expected to
be in the order of $-4.6$ to $-3.0$ MeV~fm$^5$ \cite{Ko14} as discussed in
Sect.~3. 

\begin{table}[h]
\renewcommand{\arraystretch}{1.2}
\centering
\caption{
Partial-wave contributions to the Scheerbaum factor $S_\Lambda$ (in MeV~fm$^5$). 
Same description of interactions as in Table \ref{tab:La}.
}
\begin{tabular}{|c|rrrcrrr|r|}
\hline
& $^3P_0$ & $^3D_1$ & $^3P_1$ & $^1P_1$$\leftrightarrow$$^3P_1$ & $^3P_2$ & $^3D_2$ & $^3D_3$ & Total \\
\hline
LO (550)        &    8.7 &   $-$0.2 &   $-$5.2 &    0.0 &   $-$0.9 &    0.4 &   $-$0.1 &    2.7  \\
LO (600)        &    9.3 &   $-$0.2 &   $-$5.0 &    0.0 &   $-$1.1 &    0.4 &   $-$0.1 &    3.3  \\
LO (650)        &    9.8 &   $-$0.2 &   $-$4.8 &    0.0 &   $-$1.2 &    0.4 &   $-$0.1 &    3.8  \\
LO (700)        &   10.4 &   $-$0.2 &   $-$4.7 &    0.0 &   $-$1.4 &    0.4 &   $-$0.1 &    4.4  \\
\hline
NLO (500)       &   $-$5.9 &   $-$0.6 &   $-$3.3 &    8.9 &   $-$3.4 &    0.4 &    0.2 &   $-$3.7 \\
NLO (550)       &   $-$5.2 &   $-$0.6 &   $-$2.8 &    7.5 &   $-$3.1 &    0.4 &    0.2 &   $-$3.7 \\
NLO (600)       &   $-$4.8 &   $-$0.6 &   $-$2.4 &    6.6 &   $-$3.0 &    0.4 &    0.2 &   $-$3.7 \\
NLO (650)       &   $-$4.4 &   $-$0.6 &   $-$2.0 &    5.8 &   $-$3.0 &    0.4 &    0.2 &   $-$3.7 \\
\hline
NLO$^\dagger$(650) &   $-$4.4 &   $-$0.6 &   $-$4.6 &    0.0 &   $-$3.0 &    0.4 &    0.2 &  $-$12.0 \\
NLO (600$^*$)      &    0.5 &   $-$0.1 &   $-$3.2 &    0.0 &    1.7 &    0.5 &    0.1 &   $-$0.5 \\
J\"ulich '04  & $ 4.0$ & $0.5$ & $-1.3$ & $ 4.6$ & $-9.2$  & $ 0.6$ & $-1.0$ & $-1.7$  \\
J\"ulich '94  & $-3.3$ & $1.2$ & $-3.8$ & $ 8.4$ & $-2.5$  & $ 0.6$ & $-1.3$ & $-0.4$  \\
{NSC97f}      & $-2.2$ & $0.2$ & $-12.7$ & $ 2.3$ & $-2.6$ & $-1.0$ & $-1.9$ & $-15.6$  \\
\hline
\end{tabular}
\label{tab:S}
\renewcommand{\arraystretch}{1.0}
\end{table}

In Ref.~\cite{Ko10}
the Nijmegen NSC97f potential \cite{Rij99} and a $YN$ interaction derived within the
constituent quark-model (fss2) \cite{Fu07} were found to predict $S_\Lambda = -15.4$ 
and $ -12.2$ MeV~fm$^5$ for symmetric nuclear matter at $k_F=1.35$ fm$^{-1}$,   
based on a calculation where the original potentials were represented 
by low-momentum equivalent interactions and with the so-called continuous choice for the 
intermediate spectrum. Interestingly, the LO chiral EFT potential (the one with cutoff 
mass 600 MeV), considered also in that work, yielded with $+4.8$ not only a smaller 
value but even an opposite sign for the $\Lambda$ s.p. spin-orbit potential. 
Results for the Nijmegen NSC97f potential can be also found in Ref.~\cite{Rij99}, 
however, for $k_F=1.0$ fm$^{-1}$ (and based on a different definition of $S_\La$).
Our own results, using the LO EFT potential directly, are roughly in line with those 
of Ref.~\cite{Ko10}, cf. the last column in Table~\ref{tab:S}. 
Note that
differences in the treatment of the intermediate spectrum in the propagator (gap versus 
continuous choice) do not influence the spin-dependent parts of the $G$-matrix 
interaction very much as demonstrated in Ref.~\cite{Rij99}. In particular, since 
the contributions from $P$-states and higher partial waves are rather insensitive to 
the treatment of the intermediate spectrum, the same is the case for the Scheerbaum 
factor, cf. the pertinent results for NSC97f in Refs.~\cite{Rij99} and \cite{Rij06}.
The $S$-states are more strongly affected by the intermediate spectrum and it is 
well-known that the continuous choice leads to a more attractive $G$-matrix interaction 
resulting in a 10--20~\% increase of $U_\La$ at $p_\La=0$ as compared to the gap choice 
(see, e.g., Ref.~\cite{Rij99}).

For the NLO potential we obtain $S_\La = -3.7$ MeV~fm$^5$, as aimed at in
the fitting procedure. 
It is interesting to compare the new NLO results with the ones based on the original NLO
interaction \cite{Hai13}, i.e. the one without antisymmetric spin-orbit force. Corresponding 
predictions, exemplary for $\Lambda = 650$ MeV, are included also in Table~\ref{tab:S} 
and denoted by NLO$^\dagger$(650). Clearly in this case $S_\Lambda$ is significantly larger.
A closer inspection of the respective partial-wave contributions reveals what one expects anyway,
namely that the main difference is due to the antisymmetric spin-orbit force 
whose contribution is given in the column labelled with $^1P_1$$\leftrightarrow$$^3P_1$.
It is zero for the LO interaction (because there is no antisymmetric spin-orbit force
at that order) and also for the original NLO interaction, where it has been assumed
to be zero. With the corresponding contact term included its strength can be used to
counterbalance the sizeable spin-orbit force generated by the basic interaction
so that the small $S_\Lambda$ is then achieved by a cancellation between the 
spin-orbit and antisymmetric spin-orbit components of the $G$-matrix interaction.  
 
Note, however, that the results in Table~\ref{tab:S} indicate that such a cancellation 
is not the only mechanism that can provide a small spin-orbit force. 
For example, let us look at the predictions for the interaction (denoted by) NLO (600$^*$),
where all two-meson-exchange contributions involving the $\eta$- and/or $K$ meson 
have been omitted. Here, a very small $S_\Lambda$ is achieved without any antisymmetric 
spin-orbit force. It is simply due to overall more repulsive contributions, 
notably in the $^3P_0$ and $^3P_2$ partial waves.
We want to emphasize that the interaction NLO (600$^*$) reproduces all $\La N$ and
$\Si N$ scattering data with the same high quality as the other EFT interactions 
at NLO, see the results in Ref.~\cite{Hai13}. 
It is worthwhile to mention that also the J\"ulich meson-exchange potentials
predict rather small values for the Scheerbaum factor. For both interactions there is
a substantial contribution from the antisymmetric spin-orbit force which is primarily
due to vector-meson ($\omega$) exchange \cite{Reu94}. The magnitude is comparable
to the one of the chiral EFT interaction at NLO. 
Finally, we want to draw attention to a recent lattice QCD calculation dealing with the 
$\La N$ spin-orbit force \cite{Ishii14}. In this work only a weak cancellation between 
the spin-orbit and antisymmetric spin-orbit potentials is obtained. However, one has to
keep in mind that the calculation is performed in the (flavor) SU(3) limit and,
specifically, for quark masses far away from their physical values.   

The dependence of $S_\La$ on $k_F$ can be seen in Fig.~\ref{fig:U}. Obviously, except for
the LO interaction, the density dependence is fairly weak. A weak density dependence
of $S_\Lambda$ was also observed in Ref.~\cite{Fu07} in a calculation of 
the Scheerbaum factor for a $YN$ interaction based on the quark model, and likewise
for the Nijmegen $YN$ potentials \cite{Rij99}. 

Let us now come to the $\Si$ hyperon. 
In the course of constructing the NLO EFT interaction \cite{Hai13} it turned out
that the available $YN$ scattering data can be fitted equally well with an attractive 
or a repulsive interaction in the $^3S_1$ partial wave of the $I=3/2$ $\Sigma N$ channel.  
Indeed the underlying SU(3) structure as given in Table~1 of Ref.~\cite{Hai13} 
suggests that there should be some freedom in choosing the interaction in this
particular partial wave. From that table one can see that the $^3S_1$ partial wave 
in the $I=3/2$ $\Sigma N$ channel belongs to the ``isolated'' $10$ representation 
so that the corresponding contact interaction does not enter into any of the 
interactions in the other $YN$ channels.
Since a repulsive $\Sigma$-nuclear potential can only be achieved with a repulsive 
$^3S_1$ interaction in the $I=3/2$ channel, as already discussed in the Introduction, 
the freedom mentioned above can be used to control the properties of the $\Sigma$ 
in nuclear matter within the EFT approach.

The NLO interaction presented in Ref.~\cite{Hai13} produces a moderately repulsive 
$^3S_1$ phase shift as can be seen in Fig.~9 in Ref.~\cite{Hai13}. A comparable repulsion
is also predicted by the LO potential. For the latter, calculations of the $\Sigma$ 
single-particle potential have been performed \cite{Ko10} and indicated values of
$U_\Si (p_\Si=0) \approx 12$ MeV at nuclear matter saturation density. 
Results for our (LO and NLO) EFT interactions, obtained in the present calculation, 
are compiled in Table~\ref{tab:S}. First we note that again our own LO results 
are roughly in line with those reported by Kohno \cite{Ko10}, considering, of course, 
the different prescriptions used for the treatment of the intermediate spectrum 
in the propagator. The $\Sigma$ potential depth predicted at LO and NLO is of
comparable magnitude and amounts to a repulsion in the order of $15$ to $20$ MeV. 
There is a considerable cutoff dependence visible in the LO results which is,
however, noticeably reduced at the NLO level. The J\"ulich meson-exchange 
potentials, but also the Nijmegen NSC97f interaction, yield an attractive 
$\Sigma$ s.p. potential as can be seen from Table~\ref{tab:Si}. Note that the
J\"ulich '04 interaction actually predicts a repulsive $^3S_1$ partial wave 
in the $I=3/2$ $\Si N$ channel, see the phase shifts in Fig.~9 of Ref.~\cite{Hai13}.
However, it is not strong enough to overcome the attraction provided by most of
the other partial wave contributions, cf.~Table~\ref{tab:Si}. 
The J\"ulich '94 interaction produces an unphysical (quasi) bound state in the $^1S_0$ 
partial wave of the $I=1/2$ $\Sigma N$ channel, visible as a broad shoulder in the 
$\Lambda p$ cross section at around $p_{lab} \approx$ 350 MeV/c, cf. Refs.~\cite{Reu94,Hol89}. 
As a consequence of this there is a large attractive contribution of this partial
wave to $U_\Si$ and the total $U^0_\Si$ is also large and attractive. 

A recent overview by Gal \cite{Gal10} of results from various phenomenological 
analyses of data on $\Si^-$ atoms \cite{SIGat1,SIGat2} and 
$(\pi^-,K^+)\Si$ spectra \cite{Kohno04,Kohno06} gives a range of
10-50 MeV for the isoscalar part of the $\Si$ s.p. potential and 
$\simeq +80$ MeV for the isovector component. The predictions of
the EFT interactions are well in line with those values. 
Obviously, there is a large uncertainty with regard to the actual strength of 
the isoscalar part needed for describing the data. Our EFT results are more 
at the lower end of the inferred range. It should be said, however, that significantly 
larger values are difficult to achieve. As already pointed out in \cite{Hai13}, 
there is not that much freedom to accommodate more repulsion.
Specifically, a more strongly repulsive $^3S_1$ phase shift in the $I=3/2$ $\Si N$ 
channel, as suggested for example by recent lattice QCD calculations \cite{Beane12}, 
would lead to a sizeable increase in the $\Si^+ p$ cross section, in excess 
of what is required to describe the available data, and accordingly would result 
in a dramatic deterioration of the achieved $\chi^2$. For the same reason it would
be practically impossible to realize values for the (isoscalar) $\Si$ s.p. potential 
in the order of $60$~MeV, as suggested by the perturbative calculation in \cite{Ka05}. 

Interestingly, the Scheerbaum factor $S_\Si$ is of comparable magnitude for 
all interactions considered. Like in the $\Lambda$ case, the LO EFT interaction 
predicts an opposite sign for this quantity. Note that there are no experimental 
constraints on the value of $S_\Si$ so far. 

\begin{table}[h]
\renewcommand{\arraystretch}{1.2}
\centering
\caption{
Partial-wave contributions to $U_\Sigma(0)$ (in MeV). Total results for the 
isoscalar ($U^0_\Sigma$) and isovector ($U^1_\Sigma$) decomposition according 
to Eq.~(\ref{iso}) and for 
the Scheerbaum factor $S_\Sigma$ (in MeV fm$^5$) are given as well. 
Same description of interactions as in Table \ref{tab:La}.
}
\begin{tabular}{|c||rcrrcr|rr||r|}
\hline
& \multicolumn{3}{c}{Isospin $I_0=1/2$} & \multicolumn{3}{c|}{Isospin $I_0=3/2$} & & & \\
& $^1S_0$ & $^3S_1$+$^3D_1$ & $P$ & $^1S_0$ & $^3S_1$+$^3D_1$ & $P$ & $U^0_\Si$ & $U^1_\Si$ & $S_\Si$ \\
\hline
{ LO} (550) &    8.0 &  $-15.2$ &   $-2.7$ &   $-9.9$ &   49.1 &   $-0.8$ &   28.0 &   60.8 &    8.4  \\
{ LO} (600) &    8.5 &  $-16.2$ &   $-3.2$ &  $-10.0$ &   44.6 &   $-1.0$ &   22.1 &   58.1 &    9.9  \\
{ LO} (650) &    8.6 &  $-17.1$ &  $ -3.7$ &  $-10.0$ &   40.2 &   $-1.2$ &   16.2 &   56.4 &   11.4  \\
{ LO} (700) &    8.1 &  $-18.0$ &   $-4.3$ &  $ -9.8$ &   37.0 &   $-1.3$ &   11.1 &   57.0 &   13.1  \\
\hline 
{NLO} (500) &    6.9 &  $-22.5$ &    3.4 &  $-11.4$ &   39.3 &  $ -0.2$ &   14.8 &   54.6 &  $-17.9$  \\
{NLO} (550) &    6.3 &  $-23.6$ &    2.5 &  $-10.8$ &   43.8 &  $ -0.2$ &   17.3 &   64.9 &  $-16.4$  \\
{NLO} (600) &    5.0 &  $-23.3$ &    1.8 &  $-10.2$ &   42.5 &   $-0.2$ &   14.8 &   67.8 &  $-14.8$  \\
{NLO} (650) &    4.4 &  $-22.3$ &    1.3 &  $ -9.6$ &   39.1 &   $-0.3$ &   11.9 &   65.0 &  $-14.1$  \\
\hline 
{NLO}$^\dagger$(650) &    4.4 &  $-22.3$ &    1.7 &   $-9.7$ &   39.1 &  $ -0.4$ &   12.2 &   64.1 &  $ -9.3$ \\
{NLO} (600$^*$) &    6.7 &  $-25.9$ &    3.4 &  $-10.5$ &   45.2 &  $ -0.2$ &   18.0 &   68.9 &  $ -5.5$  \\
{J\"ulich '04}  &    4.2 &  $-15.1$ &   $-8.2$ &  $-12.0$ &   11.7 &  $ -2.0$ &  $-22.2$ &   38.0 &  $-13.3$  \\
{J\"ulich '94}  &  $-18.0$ &  $-19.5$ &   $-5.1$ &   $-7.7$ &  $-15.8$ &  $ -4.1$ &  $-71.4$ &   59.1 &  $-19.8$  \\
{NSC97f}        &   15.0 &  $ -8.8$ &    0.5 &  $-12.6$ &   $-6.4$ &   $-2.6$ &  $-16.1$ &  $-32.8$ &  $-15.8$  \\
\hline
\end{tabular}
\label{tab:Si}
\renewcommand{\arraystretch}{1.0}
\end{table}

%
\section{Summary}
We have investigated the in-medium properties of a hyperon-nucleon potential, 
derived within chiral effective field theory and fitted to low-energy $\La N$ and $\Si N$ 
scattering data. In particular, we have evaluated the single-particle potential for the 
$\Lambda$ and $\Sigma$ hyperons in nuclear matter in a conventional $G$-matrix 
calculation, and the Scheerbaum factor associated with the hyperon-nucleus spin-orbit interaction.
Results are presented for a leading-order interaction published in 2006 which accounts well
for the bulk properties of the $\La N$ and $\Si N$ system \cite{Po06}
and for our recent $YN$ interaction derived up to next-to-leading order in chiral EFT 
which provides an excellent description of the available $YN$ data \cite{Hai13}. 

The predictions for the $\Lambda$ single-particle potential turned out to be in good 
qualitative agreement with the empirical values inferred from hypernuclear data. 
A depth of about $-25$~MeV is found for the NLO interaction and of about $-36$~MeV for
the LO potential. The $\Sigma$-nuclear potential is found to be repulsive, in 
agreement with phenomenological information, with values around $15$--$20$~MeV. 

Empirical information suggests that the $\Lambda$-nucleus spin-orbit interaction should 
be rather weak. Therefore, we discussed also the spin-orbit interaction and,
in particular, the role of the antisymmetric spin-orbit force in the $YN$ system. 
The chiral EFT approach yields a potential that contains, besides pseudoscalar meson
exchanges ($\pi$, $K$, $\eta$), a series of contact interactions with an 
increasing number of derivatives. 
In this approach a contact term representing an antisymmetric spin-orbit force 
arises already at NLO. It induces $^1P_1$--$^3P_1$
transitions in the coupled ($I=1/2$) $\La N$--$\Si N$ system.
The low-energy constant associated with the contact term could not be pinned down 
by a fit to the existing $\La N$ and $\Si N$ scattering data as found 
in \cite{Hai13} and, therefore, it was simply put to zero in that work. 
As demonstrated in the present study, it can be fixed, however, from investigating 
the properties of the $\La$ hyperon in nuclear matter and, specifically, it can be 
utilized to achieve a weak $\Lambda$-nuclear spin-orbit potential.
In any case, and especially with regard to the spin-orbit force, 
one has to acknowledge that nuclear matter calculations can provide only 
a first glimpse on the properties of the $YN$ interaction in the medium. A stringent 
test of the properties of the EFT interactions in the medium can be only performed 
by deducing effective interactions (from the $G$-matrix in nuclear matter) and applying
them in calculations of finite hypernuclei. 

\section*{Acknowledgements}

The authors acknowledge helpful communications with Avraham Gal, Norbert Kaiser,
Michio Kohno and Yasuo Yamamoto.  
This work is supported in part by the DFG and the NSFC through
funds provided to the Sino-German CRC 110 ``Symmetries and
the Emergence of Structure in QCD'' and by the EU Integrated
Infrastructure Initiative HadronPhysics3. Part of the numerical calculations 
has been performed on the supercomputer cluster of the JSC, J\"ulich, Germany.



\begin{thebibliography}{99}

\bibitem{Wei90}
S.~Weinberg, Phys. Lett. B {\bf 251} (1990) 288.
\bibitem{Wei91}
S.~Weinberg, Nucl. Phys. B {\bf 363} (1991) 3.

\bibitem{Epelbaum:2008ga}
  E.~Epelbaum, H.~-W.~Hammer and U.-G.~Mei{\ss}ner,
  Rev.\ Mod.\ Phys.\  {\bf 81} (2009) 1773.

\bibitem{Hai13} 
  J.~Haidenbauer, S.~Petschauer, N.~Kaiser, U.-G.~Mei{\ss}ner, A.~Nogga and W.~Weise,
  Nucl.\ Phys.\ A {\bf 915} (2013) 24.
%
%
\bibitem{HT06} O. Hashimoto and H. Tamura,
 Prog. Part. Nucl. Phys. {\bf 57} (2006) 564.

\bibitem{Gal10} A.~Gal,
  Prog.\ Theor.\ Phys.\ Suppl.\  {\bf 186} (2010) 270.
\bibitem{Botta} E.~Botta, T.~Bressani and G.~Garbarino,
  Eur. Phys. J. A {\bf 48} (2012) 41. 
%
\bibitem{Friedman07} 
  E.~Friedman and A.~Gal,
  Phys.\ Rept.\  {\bf 452} (2007) 89.

\bibitem{SIG2}
H.~Noumi et al., Phys. Rev. Lett. {\bf 89} (2002) 072301; {\bf 90} (2003) 049902 (E).
\bibitem{SIG3}
P.K.~Saha et al., Phys. Rev. C {\bf 70} (2004) 044613.

\bibitem{Kohno04} 
  M.~Kohno, Y.~Fujiwara, Y.~Watanabe, K.~Ogata and M.~Kawai,
  Prog.\ Theor.\ Phys.\  {\bf 112} (2004) 895.
\bibitem{Kohno06}
  M.~Kohno, Y.~Fujiwara, Y.~Watanabe, K.~Ogata and M.~Kawai,
  Phys.\ Rev.\ C {\bf 74} (2006) 064613.
%
%
\bibitem{SIGat1}
C.J.~Batty, E.~Friedman and A.~Gal, Phys. Lett. B {\bf 335} (1994) 273.
\bibitem{SIGat2}
J.~Mare\v s, E.~Friedman, A.~Gal and B.K.~Jennings, Nucl. Phys. A {\bf 594} (1995) 311.
%
\bibitem{Rij99}
T.~A. Rijken, V.~G.~J. Stoks and Y.~Yamamoto, Phys. Rev. C {\bf 59} (1999) 21.

\bibitem{Vid00} 
  I.~Vida\~na, A.~Polls, A.~Ramos, M.~Hjorth-Jensen and V.~G.~J.~Stoks,
  Phys.\ Rev.\ C {\bf 61} (2000) 025802.

\bibitem{Rij06}
T.~A. Rijken and Y.~Yamamoto, Phys. Rev. C {\bf 73} (2006) 044008.

\bibitem{Rij08}
T.~A. Rijken and Y.~Yamamoto, Nucl. Phys. A {\bf 804} (2008) 51.

\bibitem{Rij10}
T.~A. Rijken, M.~M.~Nagels and Y.~Yamamoto, Prog. Theor. Phys. Suppl. {\bf 185} (2010) 14.

\bibitem{Rij10a}
Y.~Yamamoto, T. Motoba and T.~A. Rijken, Prog. Theor. Phys. Suppl. {\bf 185} (2010) 72.

\bibitem{Rij13}
H.-J. Schulze and T. Rijken, Phys. Rev. C {\bf 88} (2013) 024322.

\bibitem{Fu07} Y. Fujiwara, Y. Suzuki and C. Nakamoto,
Prog. Part. Nucl. Phys. {\bf 58} (2007) 439.
%
\bibitem{Ko00} M. Kohno, Y. Fujiwara, T. Fujita, C. Nakamoto and Y. Suzuki,
Nucl. Phys. A {\bf 674} (2000) 229.
%
%
%
\bibitem{Po06}
  H.~Polinder, J.~Haidenbauer and U.-G.~Mei{\ss}ner,
  Nucl.\ Phys.\ A {\bf 779} (2006) 244.

\bibitem{Ko10} M. Kohno, Phys. Rev. C {\bf 81} (2010) 014003. 
%
\bibitem{Hi00} E.~Hiyama, M.~Kamimura, T.~Motoba, T.~Yamada and Y.~Yamamoto,
  Phys.\ Rev.\ Lett.\  {\bf 85} (2000) 270.
\bibitem{Mill10}
  D.~J.~Millener,
  Nucl.\ Phys.\ A {\bf 835} (2010) 11.
\bibitem{Mill11}
  D.~J.~Millener,
  J.\ Phys.\ Conf.\ Ser.\  {\bf 312} (2011) 022005. 
%
\bibitem{Fu08} 
  Y.~Fujiwara, M.~Kohno and Y.~Suzuki,
  Mod.\ Phys.\ Lett.\ A {\bf 24} (2009) 1031.

\bibitem{Fu00} Y. Fujiwara, M. Kohno, T. Fujita, C. Nakamoto and Y. Suzuki,
Nucl. Phys. A {\bf 674} (2000) 493.
%
\bibitem{SCHE} R.~R. Scheerbaum, Nucl. Phys. A {\bf 257} (1976) 77.
%
%
\bibitem{KaWe05} N. Kaiser and W. Weise, Phys. Rev. C {\bf 71} (2005) 015203.

\bibitem{Ka05} N. Kaiser, Phys. Rev. C {\bf 71} (2005) 068201.

\bibitem{Ka07} N. Kaiser, Phys. Rev. C {\bf 76} (2007) 068201.

\bibitem{KaWe08} N. Kaiser and W. Weise, Nucl. Phys. A {\bf 804} (2008) 60.

\bibitem{Cam07}
  J.~Martin C\'amalich and M.~J.~Vicente Vacas,
  Phys.\ Rev.\ C {\bf 75} (2007) 035207.
%
\bibitem{Reu94} A. Reuber, K. Holinde and J. Speth,
Nucl. Phys. A {\bf 570} (1994) 543.

%
\bibitem{B1} J. Dabrowski and M.Y.M. Hassan, Phys. Rev. C {\bf 1} (1970) 1883. 

\bibitem{B2} D.M. Rote and A.R. Bodmer, Nucl. Phys. A {\bf 148} (1970) 97. 

\bibitem{Pet15} S. Petschauer et al., in preparation. 

\bibitem{Yamamoto} Y. Yamamoto, private communication. 
%
%
\bibitem{Haidenbauer:2007ra}
J. Haidenbauer, U.-G. Mei{\ss}ner, A. Nogga and H. Polinder,
Lect.\ Notes Phys.\ {\bf 724} (2007) 113.

\bibitem{Pet13}
S. Petschauer and N. Kaiser,
Nucl.\ Phys.\ A {\bf 916} (2013) 1.

\bibitem{Epe05}
E. Epelbaum, W. Gl\"ockle and U.-G. Mei{\ss}ner,
Nucl.\ Phys.\ A {\bf 747} (2005) 362.

\bibitem{Nog14} A.~Nogga,
 Few Body Syst.\  {\bf 55} (2014) 757.

%
\bibitem{Ko14} M. Kohno, private communication. 

\bibitem{Fu04} Y. Fujiwara, M. Kohno, K. Miyagawa and Y. Suzuki, 
  Phys. Rev. C {\bf 70} (2004) 047002.
%
%
\bibitem{Hai05}
  J.~Haidenbauer and U.-G.~Mei{\ss}ner,
  Phys.\ Rev.\ C {\bf 72} (2005) 044005.

\bibitem{Samma} F. Sammarruca, arXiv:0801.0879 [nucl-th]. 

\bibitem{Hu14}
  J.~Hu, E.~Hiyama and H.~Toki,
  Phys.\ Rev.\ C {\bf 90} (2014) 014309.
%
%
\bibitem{Millener88} 
  D.~J.~Millener, C.~B.~Dover and A.~Gal,
  Phys.\ Rev.\ C {\bf 38} (1988) 2700.
\bibitem{Yamamoto88} 
  Y.~Yamamoto, H.~Bando and J.~\v Zofka,
  Prog.\ Theor.\ Phys.\  {\bf 80} (1988) 757.

\bibitem{Hol89}
B.~Holzenkamp, K.~Holinde and J.~Speth, Nucl. Phys. A {\bf 500} (1989) 485.

\bibitem{Ishii14}
  N.~Ishii {\it et al.}  [HAL QCD Collaboration],
  PoS LATTICE {\bf 2013} (2014) 234.
\bibitem{Beane12}
  S.~R.~Beane {\it et al.},
  Phys. Rev. Lett. {\bf 109} (2012) 172001.

\end{thebibliography}
\end{document}